\documentclass[doublecol,
figures
]{epl2} 

\usepackage{lipsum}
\usepackage{SIunits}
\usepackage{xspace}
\usepackage{color}
\usepackage{ulem}

\newcommand{\Ts}{\ensuremath{T_\mathrm{s}}\xspace}
\newcommand{\Rnot}{\ensuremath{R_0}\xspace}
\newcommand{\Unot}{\ensuremath{U_0}\xspace}

\newcommand{\Rmax}{\ensuremath{R_\mathrm{max}}\xspace}
\newcommand{\Dwater}{\ensuremath{D_\mathrm{w}}\xspace}
\newcommand{\ximax}{\ensuremath{\xi_\mathrm{max}}\xspace}
\newcommand{\nueff}{\ensuremath{\nu_\mathrm{eff}}\xspace}
\newcommand{\Reeff}{\ensuremath{Re_\mathrm{eff}}\xspace}
\newcommand{\dvisc}{\ensuremath{\delta_\textrm{$\nu$}}\xspace}
\newcommand{\dfrz}{\ensuremath{\delta_\textrm{f}}\xspace}

\renewcommand{\revision}[1]{#1}
\renewcommand{\etal}{\textit{et~al.}}

\title{Freezing-damped impact of a water drop 
}
\shorttitle{Freezing-damped drop impacts} 

\author{Virgile Thi\'evenaz\inst{1,2} \and Thomas S\'eon\inst{1} \and Christophe Josserand\inst{2}}
\shortauthor{V.~Thi\'evenaz \etal}

\institute{                    
    \inst{1} Sorbonne Universit\'es, CNRS, UMR 7190, Institut Jean Le Rond d'Alembert, F-75005 Paris, France\\
    \inst{2} Laboratoire d'Hydrodynamique (LadHyX), UMR 7646, CNRS-\'Ecole Polytechnique, 91128 Palaiseau CEDEX, 
    France
}
\pacs{nn.mm.xx}{First pacs description}
\pacs{nn.mm.xx}{Second pacs description}
\pacs{nn.mm.xx}{Third pacs description}

\abstract{
We experimentally investigate the effect of freezing on the spreading of a water drop. 
Whenever a water drop impacts a cold surface, whose temperature is lower than 0\celsius,
a thin layer of ice grows during the spreading. 
This freezing has a notable effect on the impact: at given Reynolds and Weber numbers, 
we show that lowering the surface temperature reduces the drop maximal extent. 
Using an analogy between this ice layer and the viscous boundary layer, which also grows during the spreading, 
we are able to model the effect of freezing as an effective viscosity. 
The scaling laws designed for viscous drop impact can therefore be applied to such a solidification problem, 
avoiding the recourse to a full and complex modelling of the thermal dynamics.
}

\begin{document}

\maketitle

\section{Introduction}

When a water drop impacts a cold solid surface, it spreads under its own inertia and can freeze simultaneously.
The solidification eventually yields a splat of ice, whose various possible shapes are set by the competition
between the flow and the freezing\cite{Thievenaz2020}. 
%
Freezing rain\cite{Jones1998} and aircraft icing\cite{Baumert2018}
are typical natural instances of this phenomenon which have tremendous human and economic consequences.
From an industrial point of view, mastering the simultaneous dynamics of spreading and solidification 
of an impacting drop enables precise material deposition such as
spray coating\cite{Dykhuizen1994,Pasandideh-Fard2002} and 3-D printing\cite{Lewandowski2016}.

The sole question of how does a liquid drop spread on a solid surface has given rise 
to a wide literature\cite{Josserand2016}.
Particularly, researchers have been interested in scaling laws for the maximal spreading of the drop,
depending on whether the resisting force to spreading is viscosity~\cite{Madejski1976} or 
capillarity~\cite{Clanet2004} and often both~\cite{Laan2014}.
The coupling between spreading and solidification has also been investigated, 
but mainly in the context of low impact velocity, 
namely drop deposition~\cite{Schiaffino1997a,Schiaffino97b,Tavakoli2014,Ruiter2018},
and often focusing on molten metal droplets\cite{Aziz2000,Gielen2020} \revision{or hexadecane droplets\cite{DeRuiter2017,Kant2020}}. 
\revision{Concerning water drops, previous studies have mostly been conducted on supercooled droplet impacts\cite{Schremb2018,Zhang2020} and on whether the frozen drop delaminates\cite{Ruiter2018} or cracks\cite{Ghabache2016}}.
%
But, in the general case, the maximum spreading of the solidifying impacting drop remains a largely open question.

In this letter, we investigate experimentally the spreading and freezing of water drops impacting
at room temperature on a silicon wafer cooled down below 0\celsius.
After a short description of the experimental apparatus and of qualitative observations,
we show that freezing during the impact reduces the drop maximal spreading.
Then, by observing the similarities between viscous damping and solidification, 
we are able to reduce this complex problem, which couples flow and thermal diffusion, 
to the simple case of an isothermal viscous drop impact by introducing an effective viscosity.
We thus obtain a way to express the maximal spreading radius of the drop as a function 
of the freezing rate and the liquid parameters. 

\begin{figure*}[t]
    \centering
    \includegraphics{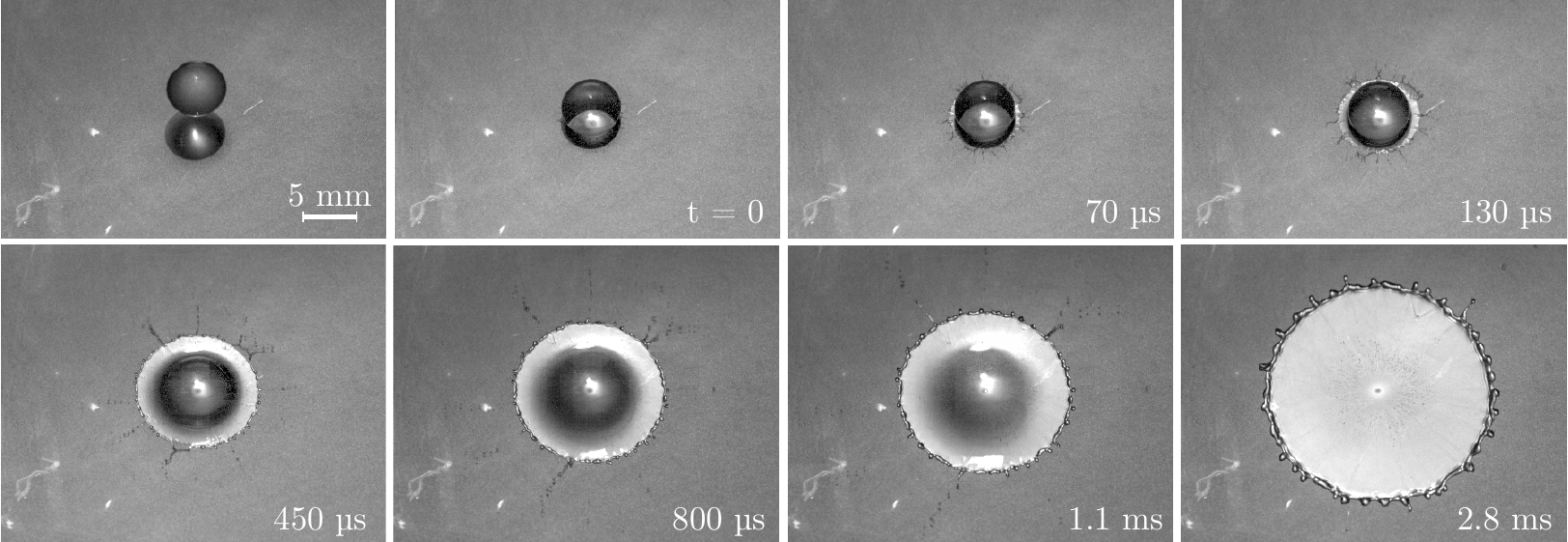}
    \caption{Image sequence of a drop impacting at $\Unot=\unit{3.1}\metre\per\second$ 
        and spreading on a silicon wafer at \unit{-19}\celsius.
    }
    \label{fig:frise}
\end{figure*}

\section{Experimental methods}
We use a syringe pump and a capillary tube to create a water drop of radius \Rnot.
This drop falls down a certain height $H$ and impacts a silicon wafer set upon a steel thermostat,
which is cooled down using liquid nitrogen. 
The surface temperature of the silicon wafer \Ts\ is measured before each impact, 
and ranges from \unit{18}\celsius\ to \unit{-109}\celsius.
By neglecting the effect of the drag onto the drop fall,   
the drop impact velocity can be estimated as $\Unot = \sqrt{2gH}$, with $g$ the gravitational acceleration.
In our experiments, \Unot ranges from \unit{2.6} to \unit{8.0}\metre\per\second.
\revision{The water drop is initially at \unit{20}\celsius, contrary to the aforementioned studies focusing on supercooled drop impact\cite{Schremb2018,Zhang2020}.}
The impact dynamic is recorded using a high-speed camera and a macro lens, 
at a rate of 75,000 frames per second which enables the precise measurement 
of the liquid film radius versus time.
The drop radius \Rnot is measured on the video for each experiment, and has an average value of \unit{1.9}\milli\metre.
In terms of dimensionless numbers, the Reynolds number $Re = 2\rho\Rnot\Unot/\mu$ varies between 7860 and 33000
and the Weber number $We = 2\rho\Rnot\Unot^2/\gamma$ between 282 and 3638.
\revision{The relevant physical quantities are taken at $20^\circ$C}: $\mu = \unit{10^{-3}}\milli\pascal.\second$ is the dynamic viscosity of water, 
$\rho = \unit{10^3}\kilo\gram\per\meter\cubed$ its density and
$\gamma = \unit{73} \milli\newton\per\metre$ its surface tension.
\revision{$\nu=\mu/\rho=\unit{1}\milli\metre\squared\per\second$ is water kinematic viscosity.}

\section{Qualitative observation}
Figure~\ref{fig:frise} presents the typical impact dynamics which occur within a few milliseconds,
for a drop impacting at $\Unot = \unit{3.1}\metre\per\second$ a silicon wafer at $\Ts = \unit{-19}\celsius$.
The drop impacts at $t=0$ then starts spreading. 
It reaches its maximal radius \Rmax after \unit{2.8}\milli\second, \revision{corresponding to the last picture of the timeline.}
During the spreading, the liquid is pushed outward into a rim, which may destabilize and form corrugations,
\revision{ and then relaxes into capillary waves.
    Simultaneously, a flat layer of ice grows from the substrate upward, and after typically one second, the drop is completely frozen.\
    It forms a frozen splat of radius \Rmax, according to the mechanisms described in Refs~\cite{Thievenaz2019,Thievenaz2020}.}

Splashing, that is the detachment of droplets from the main drop, can be observed in some cases, 
and seems in fact to be dependent on the surface temperature.
However, since the volume of liquid ejected remains negligible, we choose to neglect its effect 
on the maximum radius, and we postpone the detailed study of the splashing dynamics to future works.

\section{Maximum spreading}

\begin{figure*}[t]
    \centering
    \includegraphics[width=\hsize]{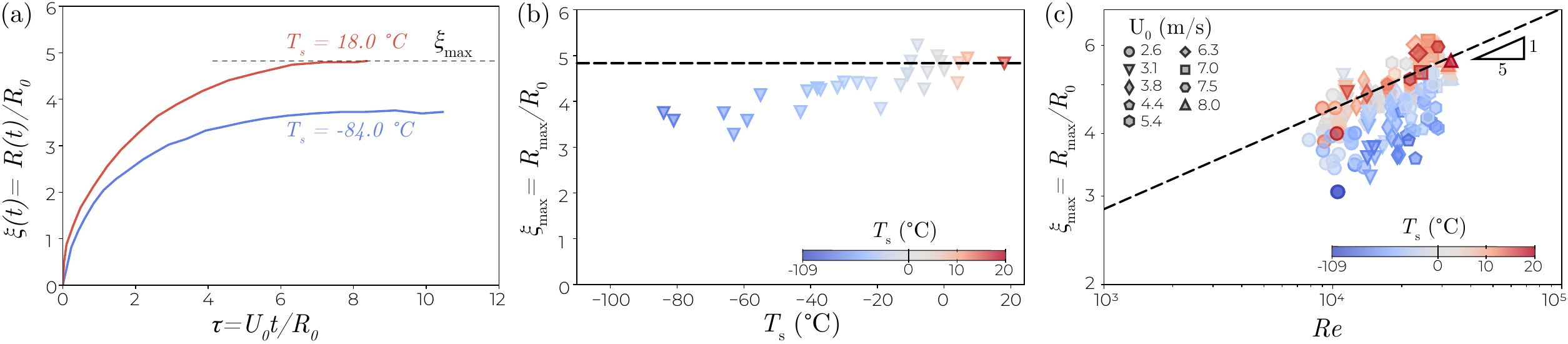}
    \caption{(a) Comparison of the non-dimensional spreading dynamic $\xi = f(\tau)$ against 
        the substrate temperature \Ts. 
        After a time of the order of a few $R_0/U_0$, the drop reaches its maximal spreading \ximax.
        At \unit{-84}\celsius\ (blue) the drop spreads less far and slower than at 18\celsius\ (red).
        (b) Maximal spreading \ximax versus the substrate temperature \Ts, 
        for a given $U_0 = \unit{3.1}\metre\per\second$.
        \ximax does not vary much above 0\celsius.
        However, below 0\celsius\ \ximax slowly drops down.
        At \unit{-100\celsius}, it has dropped 40\% of its room temperature value (dashed line).
        (c) Maximal spreading versus the Reynolds number for different substrate temperatures
        (colours, red for hot) and for $U_0$ varying between \unit{2.6} and \unit{8}\metre\per\second (symbols).
        Experiments at $\Ts \ge 0\celsius$ (warm colours) show a good agreement with a power law $1/5$
        (dashed line).
        However, this is not the case for experiments at $\Ts < 0\celsius$ (cold colours) in which water freezes.
        \revision{We chose not to display the error bars because they would be smaller than the markers, and therefore clutter up the plots.}
    }
    \label{fig:qualitative}
\end{figure*}

In order to characterize the dynamics of the spreading drop, we define the spreading parameter $\xi(t)$ 
as the ratio between the radius of the liquid film at time $t$, $R(t)$, and the initial radius of the drop \Rnot. \ximax is its maximal value, yielding :
\begin{equation}
    \xi(t) = \frac{R(t)}{\Rnot};
    \quad\quad
    \ximax = \frac{\Rmax}{\Rnot}
\end{equation}
Figure~\ref{fig:qualitative}(a) shows the evolution of the spreading parameter $\xi$ versus 
the non-dimensionalized time: $\tau = \Unot t /\Rnot$, for two different experiments corresponding to
two different substrate temperatures with the same impact velocity of \unit{3.1}\metre\per\second.
It appears clearly that at lower temperature (blue curve, $\Ts=\unit{-84}\celsius$),
the drop spreads less far than at room temperature (red curve, $\Ts=\unit{18}\celsius$).
This observation is reinforced by Figure~\ref{fig:qualitative}(b), which plots the maximal spreading parameter
\ximax versus the substrate temperature, for constant impact velocity and drop radius.
Although above 0\celsius\ the value of \ximax does not show any significant trend,
below 0\celsius\ the maximal spreading parameter decreases slowly:
\revision{\ximax drops by about 40\% over the considered substrate temperature range.}
\revision{This change of behaviour across 0\celsius\ suggests solidification plays a more important role in the damping of the drop impact than the variation of water viscosity with temperature.
We develop this point in the discussion below.}

%

Figure~\ref{fig:qualitative} (a) and (b) concern a single value of \Unot. 
In order to gather the whole range of impact velocities on the same graph, 
the maximal spreading parameter is plotted in Figure~\ref{fig:qualitative}~(c) versus the Reynolds number, 
for each of our experiments at different temperatures and impact velocities. 
The data for substrate temperatures above the melting temperature, that appear in red, 
gather all on the same line of slope $1/5$ in this logarithmic diagram.
As expected, data for negative substrate temperatures fall systematically below this line,
and the colder the substrate, the smaller the spreading radius.
The following is dedicated to propose a model to explain this discrepancy,
in other words to understand the influence of temperature on the maximum spreading of a drop.



\section{Viscous damping characterization}
The spreading of a drop on a solid substrate has a long history~\cite{Madejski1976,Pasandideh-Fard2002,Clanet2004,Fedorchenko2005,Rois09b,Eggers2010}
that has reached a consensus only recently~\cite{Laan2014,Josserand2016,Laan2016}. 
The spreading dynamic is controlled by the balance between the drop inertia on one hand, 
and both capillarity and viscous forces on the other hand.
The difficulty lies in the correct estimate of the interplay between the viscous and the capillary damping mechanisms,
and the solution is in fact deduced by analysing their asymptotic behaviours\cite{Eggers2010}. 
When viscous dissipation can be neglected,
the energy balance between inertia and capillarity gives a maximum spreading radius scaling like $\xi_\text{max} \sim We^{1/2}$. 
Oppositely, when capillary forces can be neglected, the maximum spreading radius \Rmax is reached when the thickness of the spreading drop ($\sim R_0^3/R_\text{max}^2$ using \revision{volume} conservation) becomes comparable to the thickness of the \revision{viscous} boundary layer, which obeys the usual diffusive growth law: $\dvisc=\sqrt{\nu R_\text{max}/U_0}$. This gives the scaling~\cite{Madejski1976,Clanet2004,Fedorchenko2005} \revision{:} 
\begin{equation}
\xi_\text{max} \sim Re^{1/5}
    \label{eq:Ximax}
\end{equation}
which has been verified experimentally~\cite{Lagubeau2012}.
%
These two asymptotic regimes have suggested the following general law\cite{Eggers2010}:
$\xi_\text{max}= Re^{1/5} f(P),$
where $P = We Re^{-2/5}$ is the impact parameter.
The function $f$ behaves as $f(x) \sim 1$ when $x \rightarrow \infty$ and
$f(x) \sim \sqrt{x}$ when $x \ll 1$, in agreement with the asymptotic regimes. 
Using a Pad\'e approximant method, Laan\textit{ et al.}\cite{Laan2014} have proposed the following approximation for 
$f(x)=x/(A+x)$ 
leading to the formula for the maximum spreading factor:
\begin{equation}
    \ximax \sim Re^{1/5} \frac{P^{1/2}}{A + P^{1/2}},
    \label{eq:pade}
\end{equation}
where $A$ is a fitting parameter of order one (found to be $A=1.24$ in \cite{Laan2014}). 
This formula gathers most of the known experimental and numerical results on a single curve 
and a more refined version of it has been obtained later accounting for the substrate contact angle~\cite{Laan2016}. 
Remarkably, in our experiments the impact factor $P$ varies from $7.8$ to almost $57$, 
indicating that we are in the large $P$ regime where the maximum spreading radius is controlled 
by the viscous dissipation and follows a $Re^{1/5}$ law as observed on Fig.~\ref{fig:qualitative}~(c) 
for the non-freezing case. 
This is justified by the small values of the capillary number that compares the capillary and viscous forces, 
$Ca = \mu\Unot/\gamma$, which in our experiments ranges from $0.036$ to $0.11$.
\revision{This is why we plot for comparison the $Re^{1/5}$ scaling law (dashed line) in Fig.~\ref{fig:qualitative}~(c), with its prefactor fitted to the experiments with $\Ts>0\celsius$}.

\section{Ice layer characterization}
Experiments above 0\celsius, in shades of orange in Fig.~\ref{fig:qualitative}~(c), show a good agreement with the scaling law.
On the other hand, for $\Ts<0$ the spreading parameter is systematically below this curve, 
showing that some additional damping is at work.
The physical idea behind our approach comes from the classical analogy between thermal diffusion, 
which controls the ice layer growth, and viscous diffusion\revision{,} that damps the spreading.
When the substrate is at a temperature below the freezing point, a layer of ice grows upward from the substrate, 
and its thickness is controlled by thermal diffusion.
The growth of the ice layer \dfrz is similar to that of the viscous boundary layer, namely $\dfrz = \sqrt{\alpha t}$,
where \revision{the freezing rate} $\alpha$ is a diffusion coefficient that depends on the thermal properties of the ice and 
the substrate.

This diffusive mechanism is classic for one-dimensional solidification of a liquid, 
a system known as the Stefan problem\cite{Rubinstein1971}.
In a previous article\cite{Thievenaz2019}, we wrote a variant of the Stefan problem \revision{to take heat transfers within the substrate into account, and to compute $\alpha$ quantitatively. We recall the main hypotheses: 
we consider the one-dimensional three-phase problem where a layer of ice lies between a semi-infinite solid substrate and semi-infinite water. The temperature of water is assumed constant, equal to \unit{0}\celsius. In the substrate and the ice, the temperature obeys the 1-D heat equation. By writing the continuity of temperature and heat fluxes at the ice-substrate interface, and the discontinuity of heat fluxes due to latent heat at the ice-water interface, we obtain an implicit relation between the freezing rate $\alpha$, the substrate temperature $T_\text{s}$, and the thermal parameters of the ice and substrate. This equation can be solved numerically to obtain $\alpha$.  $\alpha$ increases when the substrate temperature decreases below \unit{0}\celsius.
For more details on this model and its conclusions, see Ref.~\cite{Thievenaz2019}.}
%

For the considered temperature range, 
$\alpha$ varies from 0 (for $\Ts \ge T_m$) to \unit{1.04}\milli\metre\squared\per\second\ 
(for $\Ts=\unit{-109}\celsius$),
of the same order as the kinematic viscosity of water, $\nu=\unit{1}\milli\metre\squared\per\second$. 
Thus, the ice and the viscous layers can be considered as boundary layers to the inviscid, ice-free flow.
In other words, it means that the P\'eclet number defined as $Pe=2U_0 R_0/\alpha$ is very large 
when the Reynolds number is as well.
To get a more quantitative insight, we now consider the derivation developed 
in Refs~\cite{Rois09b,Eggers2010} and add up the ice layer dynamics.

\begin{figure}[t]
    \centering
    \includegraphics[width=.99\linewidth]{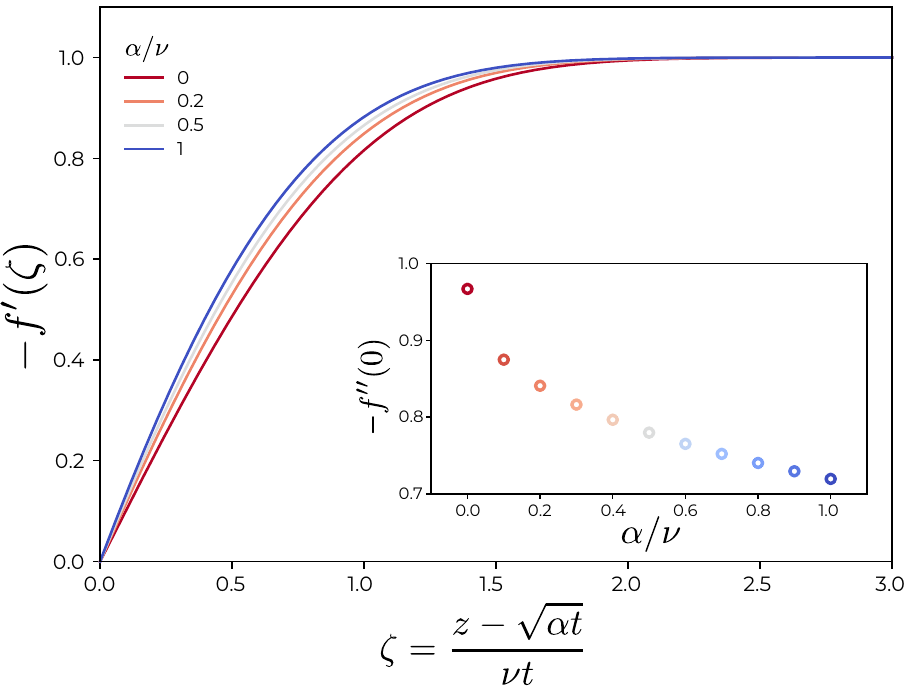}
    \caption{
        The derivative $-f'(\zeta)$ of the solution of self-similar equation \ref{bound_equ},
        for values of the ratio $\alpha/\nu$ increasing from $0$ to $1$ 
        (curves from right to left, red to blue). 
        The boundary layer thickness is defined using the slope of this curve at the origin,
        and the inset shows the evolution of this slope ($-f''(0)$, actually) 
        as the ratio $\alpha/\nu$ increases.
        }
    \label{fig:visc_eff}
\end{figure}

\section{Ice-viscous boundary layer} 
We start with the r-component of the axisymmetric Navier-Stokes equation for incompressible flows 
%
with the boundary condition $v_r=0$ for $z=\sqrt{\alpha t}$, indicating that the ice layer grows homogeneously from the substrate. 
Using the classical argument of Prandtl's boundary layer theory~\cite{Prandtl}, we can show that the Navier-Stokes equation 
reduces to: 
\begin{equation}
\partial_t v_r + v_r\partial_r v_r + v_z\partial_z v_r = \nu\partial_z^2 v_r,
\label{bound}
\end{equation}
and it is more convenient to use the stream-function $\psi$ to satisfy the boundary condition, yielding:
\begin{equation}
v_r = -\frac{\partial_z \psi}{r}, \qquad v_z = \frac{\partial_r \psi}{r}.
\label{psi}
\end{equation}
In the inviscid case, we have $\psi = -r^2z/t$ and since both the viscous boundary layer $\dvisc$ and the ice layer thickness $\dfrz$ 
follow the same square-root-in-time scaling, it suggests the ansatz:

\begin{equation}
\psi = \sqrt{\nu}\frac{r^2}{\sqrt{t}} f\left(\frac{z-\sqrt{\alpha t}}{\sqrt{\nu t}}\right).
\label{psi_nu}
\end{equation}
For $\alpha=0$ and $f(\zeta) = -\zeta$, taking $\zeta = \frac{z}{\sqrt{\nu t}}$ as the self-similar variable,
the inviscid result is recovered. 
Inserting (\ref{psi_nu}) into the boundary layer equation (\ref{bound}),
we find 
\begin{equation}
f' + \left(\eta+\sqrt{\frac{\alpha}{\nu}} \right) f''/2 + f'^2 - 2ff'' = -f''',
\label{bound_equ}
\end{equation}
with the boundary conditions imposing zero velocity at the ice-water interface, and the inviscid horizontal velocity field far from the plate:
\begin{equation}
f'(\infty) = -1, \qquad f(0) = 0, \qquad f'(0) = 0.
\label{bound_b}
\end{equation}
This self-similar equation shows that the structure of the boundary layer in the fluid is influenced 
by the growth of the ice layer through the single additional term $\sqrt{\alpha/\nu}f''/2$ 
which obviously vanishes when the ice layer is absent ($\alpha=0$).
Equation~(\ref{bound_equ}) can be solved numerically using a shooting method~\cite{Eggers2010},
and the main plot of Fig. \ref{fig:visc_eff} 
shows the variations of the solution's derivative ($-f'$)
as the ratio $\alpha/\nu$ varies from $0$ (non-freezing case) to $1$ 
(about the maximum value of the ratio in our experiments). 
Since $v_r=-rf'(\zeta)/t$, this derivative provides a good description of the velocity field in the boundary layer.
Indeed, in the boundary layer the horizontal velocity starts from zero at the ice-water interface (no-slip), 
and reaches the inviscid flow behaviour when $-f'(\zeta) \sim 1$.
Figure~\ref{fig:visc_eff} 
reveals little change in the boundary layer as $\alpha/\nu$ varies between $0$ and $1$.
Finally, the boundary layer thickness $\dvisc$ can be defined directly from this solution, estimating the length needed for the horizontal velocity to approach the inviscid behavior yielding:
\begin{equation}
    \dvisc= \frac{\sqrt{\nu t}}{-f''(0)}
\end{equation}
The inset of Fig. \ref{fig:visc_eff} 
plots $-f''(0)$ versus the ratio $\alpha/\nu$, 
and reveals that the boundary layer prefactor $-1/f''(0)$ increases at most $25 \%$ over our experimental range.
Therefore, a first approximation of the viscous boundary layer can be taken as $\dvisc=\sqrt{\nu t}$.
In conclusion, this result shows that the viscous boundary layer grows on top of the freezing layer, 
with a negligible coupling between them.
This configuration can be globally considered as a mixed boundary layer of size $\delta=\dvisc+\dfrz$.





\section{Effective viscosity}


\begin{figure}[t]
    \centering
    \includegraphics[width=.99\linewidth]{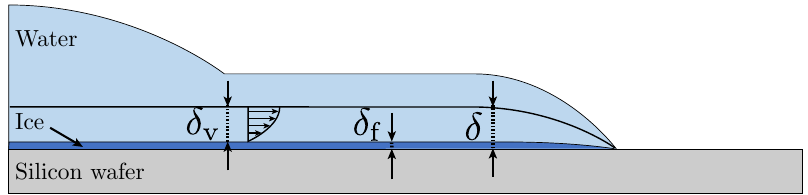}
    \caption{
        Schematic of a drop spreading and freezing on a cold substrate.
        Viscous damping occurs across the viscous boundary layer $\dvisc = \sqrt{\nu t}$,
        whereas the growing ice layer has thickness $\dfrz = \sqrt{\alpha t}$.
        As a result, the spreading is damped across a total thickness $\delta = \dvisc + \dfrz$.
        }
    \label{fig:visc_effb}
\end{figure}

Within this framework, Figure~\ref{fig:visc_effb} 
shows a schematic cross-section 
of the drop spreading and freezing.
As the liquid spreads, the viscous boundary layer of thickness $\dvisc = \sqrt{\nu t}$
grows from the solid surface upwards, and concentrates the viscous dissipation.
At the same time, the ice layer $\dfrz = \sqrt{\alpha t}$ grows with a similar square-root-of-time dynamic.
When the liquid freezes, it stops moving, which constitutes a loss of kinetic energy.
Hence, viscous dissipation and freezing both reduce the drop inertia.
The dynamics of the two phenomena are associated and damping operates across the thickness of the mixed boundary layer $\delta$:
\begin{equation}
    \delta  = \sqrt{\nu t} + \sqrt{\alpha t}.
\end{equation}
If we now assume that the spreading stops when the mixed boundary layer $\delta$ reaches the free surface, 
we obtain a similar result as that of the non-freezing case (Eq. (\ref{eq:Ximax})), by defining the effective viscosity \nueff such as $\delta = \sqrt{\nueff t}$:
\begin{equation}
    \nueff = \nu \left( 1+ \sqrt\frac{\alpha}{\nu} \right)^2.
    \label{eq:nueff}
\end{equation}
Using \nueff, we can define the effective Reynolds number \Reeff:
\begin{equation}
    \Reeff = \frac{Re}{\left(1+ \sqrt\frac{\alpha}{\nu} \right)^2}.
    \label{eq:Reeff}
\end{equation}
Therefore, we can replace $Re$ by \Reeff in Eq. (\ref{eq:Ximax}), and in the viscosity-dominated regime considered in this paper, we expect the following spreading law:
\begin{equation}
    \ximax \sim \Reeff^{1/5}
    \label{eq:scalingXimax}
\end{equation}

\revision{In fact, Eq. (\ref{eq:Ximax}) uses the drop volume conservation and a correction should be added because of the $10\%$ volume expansion of the ice by comparison with water. However, we neglect this correction concerning only the solid part of the drop, leading at most to a $2\%$ correction for $\ximax$ that is below the typical error of our measurements.}

Above 0\celsius, $\alpha=0$ and thus $\Reeff=Re$, which returns the viscous scaling $\ximax=Re^{1/5}$.
Below 0\celsius, the colder the temperature the higher the freezing rate $\alpha$, 
the smaller the effective Reynolds number.

\begin{figure}[t]
    \centering
    \includegraphics{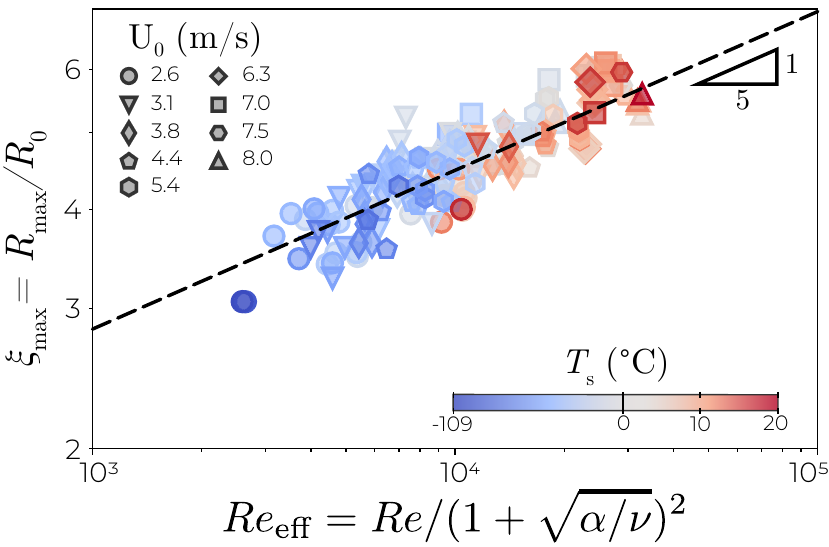}
    \caption{
        Maximal spreading versus the effective Reynolds number (Eq.~\ref{eq:Reeff}).
        The new definition of the Reynolds number shifts the experiments where freezing occurs (blue) to the left,
        and yields a good agreement with the power law $1/5$.
        The dashed line has a slope $1/5$ and its prefactor is fitted to the experiments above 0\celsius\ only.
    }
    \label{fig:rescaled}
\end{figure}

With the aim of confronting this model against our experimental data,
Figure~\ref{fig:rescaled} plots the maximal spreading parameter versus the effective Reynolds number.
As in Figure~\ref{fig:qualitative}~(c), the dashed line represents the power law $\ximax \sim \Reeff^{1/5}$ fitted
with the experiments above 0\celsius, which remains unchanged. 
Only the data representing the experiments below 0\celsius\ have shifted to the left towards the small Reynolds numbers.
There is now a good agreement between all data and the power law, for the whole range of temperature.
This proves that the freezing of the spreading drop can be appropriately modelled through an effective viscosity,
whose magnitude is set by the freezing rate.

\section{Discussion}



A notable point of Figure~\ref{fig:rescaled} is that all experiments collapse well not only on a line of
slope 1/5 as expected, but on the same line as the experiments above 0\celsius, 
without requiring any additional fitting parameter.
This is probably due to our good estimation of the freezing rate $\alpha$\cite{Thievenaz2019}.
Moreover, the good agreement of experiments with the scaling law $\ximax = \Reeff^{1/5}$ 
suggests that the freezing and the spreading of the drop follow distinct dynamics.
Since the model we used to derive the freezing rate 
assumes the liquid remains still while freezing, 
we infer that the thermal diffusion which controls the freezing rate is apparently not affected by the flow.
\revision{Also, the proposed model completely ignores what happens at the contact line and only considers bulk viscous dissipation. 
The good agreement between this theory and the experiments shows that the mechanism which stops the spreading may not reside at the contact line, as previously thought\cite{Schiaffino1997a,Tavakoli2014,DeRuiter2017}.}


Gielen~\textit{et al.}~\cite{Gielen2020} recently conducted an analogous study with molten tin drops impacting
a sapphire substrate, and observed a similar reduce in the maximal spreading while decreasing
the substrate temperature.
Their explanation is based on the sole growing solid layer, without taking into account the viscous boundary layer.
In the present framework, it amounts to neglect the kinematic viscosity $\nu$ compared to the freezing rate $\alpha$. 
The model of Gielen~\textit{et al.} is consistent with their experiments, but did not match ours, 
because in our case the freezing rate $\alpha$ is comparable to the kinematic viscosity $\nu$, 
although it remains smaller.
On the other hand, tin has a latent heat more than five times lesser than water which makes it easier to freeze, 
and its liquid phase is four times less viscous, 
explaining why the effect of viscosity in the impact of molten tin drops can be neglected.


Rather than an effect of solidification, the decrease of \ximax with temperature could be due to the increase of viscosity of water depending on the temperature.
Indeed, between 20\celsius\ and 0\celsius, the viscosity of water almost doubles, 
from \unit{1}\milli\pascal.\second\ to \unit{1.79}\milli\pascal.\second\cite{Lide2012}.
For the flow viscosity to increase further below 0\celsius, one would need have supercooled water,
otherwise it would just be solid.
The viscosity of supercooled water does indeed increase at lower temperatures:
at -20\celsius\ it is \unit{4.33}\milli\pascal.\second, more than twice its value at 0\celsius\cite{Hallett1963}.
However, it is unlikely that the temperature dependence of viscosity be the main cause for the damping for two reasons :
firstly, we do not observe any decrease in the maximal spreading above 0\celsius, 
which could be attributed to the increase in the viscosity of water.
Secondly, supercooled water does not exist below -40\celsius.
This is contradictory with our observations which are consistent over the whole range of temperature 
from \unit{18}\celsius\ to \unit{-109}\celsius.

\revision{
    Although neglecting the increase in water viscosity due the decrease in temperature apparently conflicts with the study of viscous damping, it is easily justified by looking at the volume of water which it concerns.
    This volume of water, which is cold, thus more viscous than at room temperature, yet remains liquid, corresponds to the thermal boundary layer which grows on top of the ice. 
    Its thickness grows roughly like $\sqrt{\Dwater t}$, with $\Dwater=\unit{0.14}\milli\metre\squared\per\second$ being water heat diffusivity. 
    Since the freezing rate $\alpha$ is most of the time greater than \Dwater, it means this cold liquid layer freezes as soon as it cools down, so its volume remains negligible at all times.
    Therefore, viscous dissipation therein is negligible compared to dissipation in the bulk at \unit{20}\celsius.
}

\revision{We notably dismissed the nucleation of ice and whether water supercooling occurs in the process of freezing the spreading drop\cite{Kant2020}. During our study, we did not observe the typical evidences of supercooling, such as dendrites or patches of ice growing from different parts of the water-silicon interface. 
    Supercooling would typically happen if the substrate temperature and the initial drop temperature were closer to the melting point.
    Studies focusing on that matter (such as Ref.\cite{Kant2020}) consider the vicinity of the melting point and materials with a lower heat capacity (hexadecane), which is not in the range of our study.
    Furthermore, our observations are well explained under the assumption of a flat ice-water interface, which only makes sense in the absence of supercooling. This is coherent with our previous work\cite{Thievenaz2019,Thievenaz2020}, in which predictions assuming water was not supercooled were quantitatively verified.
For those reasons, we chose to neglect water supercooling throughout our study.} 


\revision{
Finally, although our results are limited to the visco-inertial regime ($P\gg1$), 
they suggest that the spreading law given by equation (\ref{eq:pade}) could be expanded 
to any freezing drop impact by redefining the effective impact parameter $P = We \Reeff^{-2/5}$,
leading to :
\begin{equation}
    \ximax \sim \Reeff^{1/5} \frac{P^{1/2}}{A + P^{1/2}},
    \label{eq:pade_eff}
\end{equation}
In the present regime, this expression does not provide a better fit to our data than 
the purely viscous $\Reeff^{1/5}$ law.
Its validation would require new experiments, and deserves a dedicated study.
}

\section{Conclusion}
When a water drop impacts a surface colder than its melting temperature, it freezes as it spreads.
Compared to the isothermal case the solidification reduces the maximal extent of the drop, as much as $40\%$ for a
temperature of \unit{-100}\celsius.
Since the ice layer beneath the liquid film grows similarly to the viscous boundary layer,
we showed that this effect could be modelled as an effective viscosity, which is a function of the ratio 
between the freezing rate $\alpha$ and the liquid kinematic viscosity $\nu$.
This approach matches our experiments well, and is consistent with the existing literature on the subject.
Therefore, this concept of freezing-induced effective viscosity offers an interesting tool for the study of other
complex systems involving flows and phase transitions. 


\bibliographystyle{eplbib}
\bibliography{FreezingDamped}

\end{document}